\documentclass[letter]{iopart}

\usepackage{graphicx}

\begin{document}

\letter{Adaptable-radius, time-orbiting magnetic ring trap for Bose-Einstein condensates}
\author{A. S. Arnold}
\address{Department of Physics, University of Strathclyde, Glasgow G4 0NG, UK}
\ead{a.arnold@phys.strath.ac.uk}

\begin{abstract}
We theoretically investigate an adjustable-radius magnetic storage ring for laser-cooled
and Bose-condensed atoms. Additionally, we discuss a novel time-dependent variant of this
and other ring traps. Time-orbiting ring traps provide a high optical access method for
spin-flip loss prevention near a storage ring's circular magnetic field zero. Our
scalable storage ring will allow one to probe the fundamental limits of condensate Sagnac
interferometry.
\end{abstract}

The field of atom optics \cite{erlcha} has been transformed by laser cooling
\cite{rmplascool} and Bose-Einstein condensation (BEC) \cite{rmpbec}. Prior to laser
cooling, atoms could only be magnetically deflected by small angles, due to the
relatively weak nature of the interaction and the large atomic velocities involved
\cite{hinds}. With the advent of laser cooling, atomic beams have been magnetically
deflected by much larger angles \cite{meschede}, and cold atomic clouds have been
magnetically reflected \cite{hinds}. Storage rings for cold atoms are atom-optical
elements which have only been realised very recently \cite{sau,web,prent}.

\begin{figure}[ht]
\begin{center}\mbox{\includegraphics[clip,width=1\columnwidth]{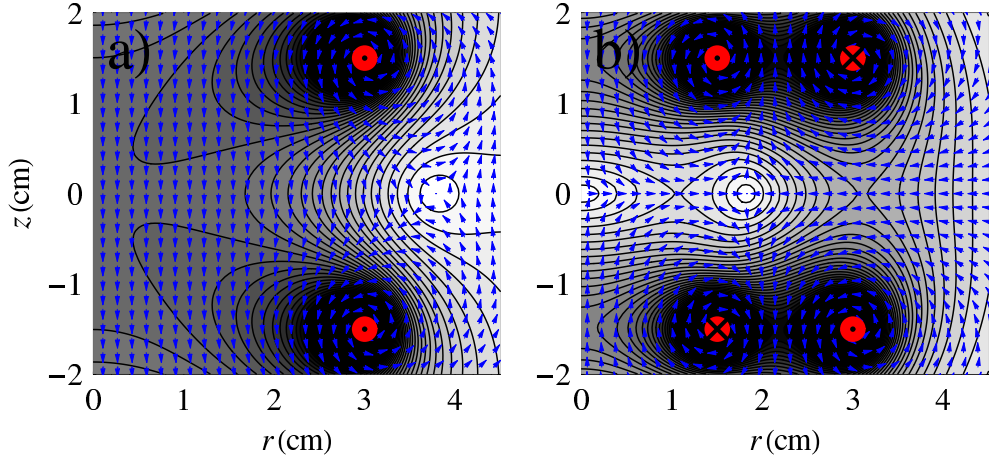}}\end{center}
\caption{Images a) and b) depict storage rings formed of 2 or 4 circular coils. Contours
(with spacing $25\,\rm{G})$ and vectors indicate the magnetic field magnitude and
direction, respectively. All coils carry a current of $1500\,\rm{A}.$ The storage ring is
formed at zeros of the magnetic field (Image a): $(r_0,\,z_0)=(3.78,\,0),$ Image b):
$(r_0,\,z_0)=(1.82,\,0)$ (the storage ring) or $(0,\,0)$). Note that the storage ring
position approaches the average coil radius if the coil separation is much smaller than
the coil radii.}
 \label{contvec}
\end{figure}

To date, all experimental cold atom storage rings are generated with magnetic fields
where the (fixed) radius of the ring is determined by the radii of the ring's magnetic
elements (Fig.~\ref{contvec}). Additionally, these storage rings are formed at a zero in
the magnetic field, thus atoms - particularly the cold atoms - can spin-flip into
non-trapped magnetic quantum states at the centre of the ring \cite{petr}. A long
current-carrying wire along the symmetry ($z$) axis of the storage ring can generate an
azimuthal magnetic field around the ring, preventing spin-flip losses \cite{jmo}.
However, optical access to the storage ring is limited by such a wire, and the wire must
either be in-vacuo or a `hole' in the vacuum chamber is required. We present a new
variable-radius magnetic storage ring. In addition we show how, by adding a time-orbiting
term to the magnetic field, it is possible to create a storage ring which is stable with
respect to spin flips, without the necessity of an axial current-carrying wire.

We note that a cold molecule storage ring based on electric dipole forces has also been
developed \cite{Mei}, however, although cold molecules are interesting in their own
right, atoms are more suitable candidates for many interferometry experiments. In
contrast to molecules, atoms can easily be prepared in the same quantum state, and they
can be cooled to temperatures hundreds of times colder than molecules due to the ease
with which they can be laser cooled.

Magnetic atom-optical elements use the Stern-Gerlach potential $U=\mu_B g_F m_F B$ an
atom experiences whilst moving adiabatically in a magnetic field of magnitude $B,$ where
$\mu_B$ is the Bohr magneton, $m_F$ is the atom's hyperfine magnetic quantum number, and
$g_F$ is the Land\'{e} g-factor. We consider weak-field-seeking atoms $(g_F m_F>0).$

We begin by discussing a second-order model for the magnetic field, as it neatly
describes the main features of the storage ring. However, this approximation is only used
as an illustration and all figures in this paper are calculated using the full
Biot-Savart law for circular coils \cite{jackson}. If a cylindrically symmetric magnetic
coil configuration is used, then Maxwell's equations and the on-axis behaviour of the
magnetic field lead to the complete 3D Taylor expansion of the magnetic field. In
particular, the second-order on-axis 1D magnetic field,
$B_z(r=0,z)=B_0+B_1\,z+B_2\,z^2/2,$ gives the complete 3D second-order magnetic field:
\begin{equation}
\textbf{B}(r,z)=\{B_r,B_z\}=\{-\frac{B_1}{2}r-\frac{B_2}{2}rz,\,B_0+B_1\,z+\frac{B_2}{2}\left(z^2-\frac{r^2}{2}\right)\}.
 \label{field}
\end{equation}
We note that these results are contrary to those of Ref.~\cite{kett}, where the behaviour
of the axial magnetic field in the $z=0$ radial plane was used to generate a 3D
second-order magnetic field different to that of Eq.~\ref{field}.

An axially-separated double trap \cite{wilson} is formed when $B_0\,B_2<0$ and a ring
trap is formed when $B_0\,B_2>0.$ By solving Eq.~\ref{field} and finding the magnetic
field zero, the radial and axial location of the storage ring are found:
 \begin{equation}
 r_0=\sqrt{4B_0/B_2-2{B_1}^2/{B_2}^2},\;\;\;\;z_0=-B_1/B_2.
 \label{zeropos}
 \end{equation}
 The gradient of the magnetic field magnitude near the centre of the storage ring is identical in the
 axial $(B_{1z})$ and radial $(B_{1r})$ directions,
 \begin{equation}
  B_{1r}=B_{1z}=\sqrt{B_0 B_2 -{B_1}^2/2},
 \end{equation}
and atoms experience a potential which is approximately conical:
 \begin{equation}
  U\propto|\textbf{B}|\approx B_{1r}\sqrt{(r-r_0)^2+(z-z_0)^2}.
 \label{pot}
 \end{equation}

\begin{figure}[ht]
\begin{center}\mbox{\includegraphics[clip,width=1\columnwidth]{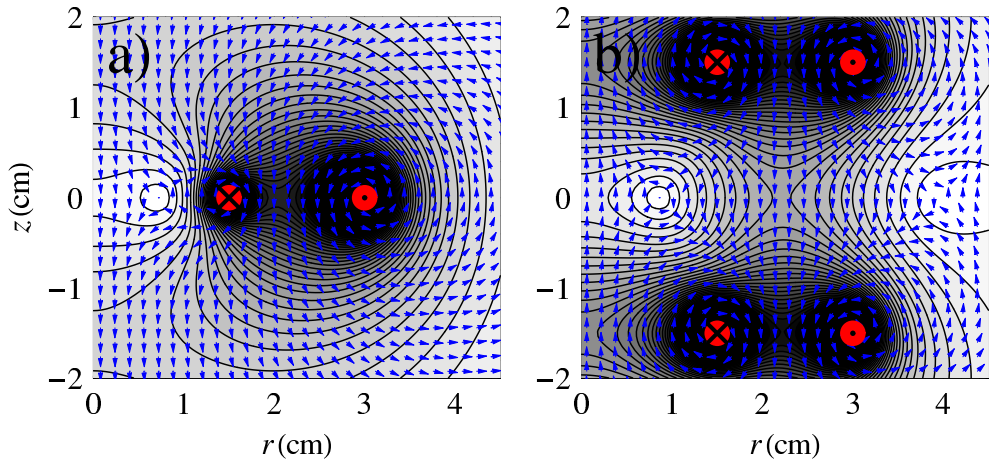}}\end{center}
\caption{Image a) depicts the simplest form of adjustable ring trap, comprised of two
co-planar circular coils. Contours and vectors indicate the magnetic field magnitude and
direction, respectively. This kind of trap would probably require the coils to be place
in-vacuo. Image b) depicts a storage ring in which each of the coils of image a) are
replaced by an axially separated pair of coils. In this case the coils could be placed
outside of the vacuum cell. Images a) and b) use the same contour spacing of
$25\,\rm{G},$ and the storage ring is formed at zeros of the magnetic field (Image a):
$(r_0,\,z_0)=(0.72,\,0),$ Image b): $(r_0,\,z_0)=(0.86,\,0)$ (the strong storage ring) or
$(3.99,\,0)$). In a) $I_2/I_1=2.33$ and in b) $I_2/I_1=0.80,$ where $I_1$ and $I_2$
correspond to the currents in the small $(r_1)$ and large $(r_2)$ radius coils,
respectively.}
 \label{contvec2}
\end{figure}

The simplest geometry for creating a ring trap is with two concentric co-planar circular
coils of radius $r_1$ and $r_2>r_1$ respectively, carrying currents with magnitudes $I_1$
and $I_2>r_2/r_1\,I_1$ in opposite directions (Fig.~\ref{contvec2} a)). The radius of the
storage ring is adjusted by varying the ratio of currents $I_2/I_1$ in the coils. Such a
storage ring is formed in the plane of the two coils, and therefore requires an in-vacuo
coil system. Two axially-displaced concentric coils can also be used to form the storage
ring. However, the symmetry and additional confinement provided by two concentric,
equally spaced, pairs of opposing `Helmholtz' coils, make them a more apt configuration
for an ex-vacuo coil system (Fig.~\ref{contvec2} b)). Note that the orientation of the
currents in the standard two and four coil storage ring (Fig.~\ref{contvec}) is different
from the equivalent variable radius storage ring (Fig.~\ref{contvec2}). The maximum coil
current used in all of our figures is $I_1=1500$ Amp-turns, which is identical to some of
the coils used in our existing storage ring experiment \cite{web,jmo}.

\begin{figure}[ht]
\begin{center}\mbox{\includegraphics[clip,width=1\columnwidth]{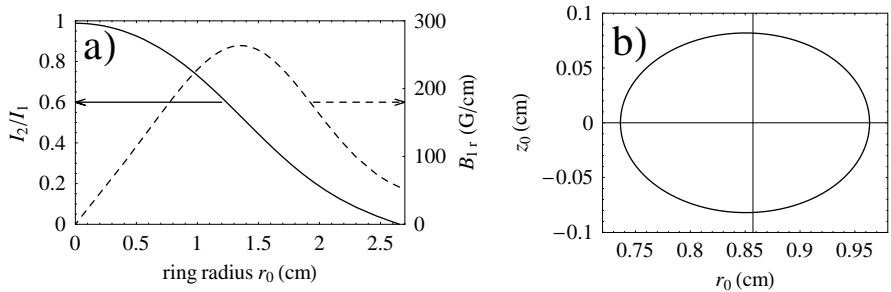}}\end{center}
\caption{Image a) shows the ratio of current $I_2/I_1$ (solid curve) between the larger
and smaller radius rings, and the gradient of the magnetic field magnitude $B_{1r}$
(dashed curve) in the storage ring as a function of the storage ring radius $r_0.$ Image
b) shows the locus of the ring of zero magnetic field when $I_2/I_1=0.8,$ $\eta_r=0.06$
and $\eta_z=0.25.$ For a 3D animation of the time-dependent behaviour of the TORT's ring
of zero magnetic field see Ref.~\cite{web}.}
 \label{TORTgraf}
\end{figure}

In order to generate a time-orbiting ring trap (TORT) we simply need to time-dependently
alter the radial and axial position of the ring of zero magnetic field (given
approximately in Eq.~\ref{zeropos}). In particular, adding small sinusoidally varying
$B_1$ and $B_0$ terms mainly shift the $z$ and $r$ positions, respectively, of the
storage ring magnetic zero point. Varying $B_1$ and $B_0$ in quadrature leads to the kind
of magnetic field zero trajectory shown in Fig.~\ref{TORTgraf} b). This effect can be
either be achieved by using separate pairs of anti-Helmholtz $(B_1)$ and Helmholtz $(B_0$
and $B_2)$ or `true' Helmholtz $(B_0)$ coils, or by adding sinusoidal modulations with
the appropriate amplitude and relative phase to the two separate coils of one of the
Helmholtz coil pairs in Fig.~\ref{contvec2} b). For the magnetic zero trajectory shown in
Fig.~\ref{TORTgraf} b), we change the currents of the `pure' Helmholtz coils at radius
$r_2=3\,\rm{cm}$ in Fig.~\ref{contvec2} b) from $I_2$ to $I_{2a}=I_2(1+\eta_r
\sin(2\pi\nu t)+\eta_z \cos(2\pi\nu t))$ and $I_{2b}=I_2(1+\eta_r \sin(2\pi\nu t)-\eta_z
\cos(2\pi\nu t)),$ where $\eta_r=0.06$ and $\eta_z=0.25$ change the $r$ and $z$ position
of the storage ring, respectively.

A TORT trap transforms a storage ring's conical magnetic potential (Eq.~\ref{pot}) into a
time-averaged hyperbolic potential. A useful parameter is the time-averaged magnet field
magnitude at trap centre $B_b.$ The frequency of the TORT current modulation
$\nu\approx10\,\rm{kHz}$ needs to be at approximately the geometric mean of the harmonic
TORT trap frequency $(\approx100\,\textrm{Hz}\;(\propto {B_b}^{-1/2}))$ and the trap's
Larmor frequency $(\nu_L\approx1\,\textrm{MHz}\;(\propto B_b))$ \cite{petr}. This
frequency is typical for time-orbiting quadrupole traps, and we note that the coils
involved in both time-orbiting quadrupole and ring traps have similar dimensions. We have
concentrated here on a time-orbiting variant of our adjustable-radius storage ring,
however similar techniques exist for making a time-orbiting fixed-radius storage ring.

In order to make a Bose condensate in the storage ring, it is necessary to localise the
atoms in a magnetic potential that is three-dimensionally harmonic for efficient
evaporative cooling. In Ref.~\cite{jmo} we discuss how we use four magnetic pinch coils
to transform a sector of a storage ring into both a magneto-optical trap \cite{raab} and
a Ioffe-Pritchard magnetic trap \cite{pri}. The same technique can be used with our
adjustable-radius storage ring. After evaporative cooling and subsequent condensation in
the Ioffe-Pritchard trap, the pinch coils can be adiabatically turned off and Bragg
scattering \cite{bragg} can be used to coherently launch condensate wavepackets around
the TORT.

An important limitation to the performance of interferometers using Bose-Einstein
condensates is that in storage rings (and other traps with elongated geometries),
condensates break up into fragments with different phases. Such phase fluctuations
\cite{phasefluc} can be reduced if the temperature in the trap is very low. The tunable
radius of the ring allows one to find ideal conditions for Sagnac interferometry: i.e. by
finding the maximum ring area over which phase fluctuations are negligible.

We believe that an adjustable-radius magnetic storage ring trap for Bose-condensed atoms
will be a very useful atom-optical tool. Additionally, our novel time-orbiting ring trap
(TORT), does not require an axial current-carrying wire, providing a high optical access
method for spin-flip loss prevention near a storage ring's circular magnetic field zero.
Our scalable storage ring will allow one to study persistent currents, and probe the
fundamental limits of Bose-Einstein condensate Sagnac interferometry.
\newline

This research was supported by the UK EPSRC, the University of Strathclyde and INTAS.


\section*{References}
\begin{thebibliography}{99}
 \bibitem{erlcha}
    Adams C S and Riis E 1997 \textit{Prog. Quant. Electr.} \textbf{21} 1
 \bibitem{rmplascool}
    Chu S 1998 \textit{Rev. Mod. Phys.} \textbf{70} 685; Cohen-Tannoudji C N 1998 \textit{Rev. Mod. Phys.} \textbf{70} 707;
    Phillips W D 1998 \textit{Rev. Mod. Phys.} \textbf{70} 721
 \bibitem{rmpbec}
    Cornell E A and Wieman C E 2002 \textit{Rev. Mod. Phys} \textbf{74} 875; Ketterle W 2002 \textit{Rev. Mod. Phys} \textbf{74} 1131
 \bibitem{hinds}
    Hinds E A and Hughes I G 1999 \textit{J. Phys. D} \textbf{32} R119
 \bibitem{meschede}
    Goepfert A \textit{et al.} 1999 \textit{Appl. Phys. B} \textbf{69} 217
 \bibitem{sau}
    Sauer J A, Barrett M D and Chapman M S 2001 \textit{Phys. Rev. Lett.} \textbf{87} 270401
 \bibitem{web}
    Arnold A S, Garvie C S and Riis E 2002,
    http://www.photonics.phys.strath.ac.uk/Research/BEC/Ring.html;
 \bibitem{prent}
    Rooijakkers W \textit{et al.} 2003, http://www.eps.org/aps/meet/DAMOP03/baps/abs/S390008.html
 \bibitem{petr}
    Petrich W \textit{et al}. 1995 \textit{Phys. Rev. Lett.} \textbf{74} 3352
 \bibitem{jmo}
    Arnold A S and Riis E 2002 \textit{J. Mod. Optics} \textbf{49} 95
 \bibitem{Mei}
    Crompvoets F M H \textit{et al.} 2001 \textit{Nature} \textbf{411} 174
 \bibitem{kett}
    Ketterle W and Pritchard D E 1992 \textit{Appl. Phys. B} \textbf{54} 403
 \bibitem{wilson}
    Thomas N R, Wilson A C and Foot C J 2002 \textit{Phys. Rev. A} \textbf{65} 063406
 \bibitem{jackson}
    Jackson J D 1999 \textit{Classical Electrodynamics} 3rd edn (New York: Wiley) p 182; Good R H 2001 \textit{Eur. J. Phys.} \textbf{22} 119
 \bibitem{raab}
    Raab E L \textit{et al}. 1987 \textit{Phys. Rev. Lett.} \textbf{59} 2631
 \bibitem{pri}
    Pritchard D E 1983 \textit{Phys. Rev. Lett.} \textbf{51} 1336
 \bibitem{bragg}
    Stenger J \textit{et al}. 1999 \textit{Phys. Rev. Lett.} \textbf{82} 4569;
    Kozuma M \textit{et al} 1999 \textit{Phys. Rev. Lett.} \textbf{82} 871
 \bibitem{phasefluc}
    Petrov D S, Shlyapnikov G V and Walraven J T M 2001 \textit{Phys. Rev. Lett.} \textbf{87} 050404;
    Dettmer S \textit{et al}. 2001 \textit{Phys. Rev. Lett.} \textbf{87} 160406;
    Hellweg D \textit{et al}. 2003 \textit{Phys. Rev. Lett.} \textbf{91} 010406;
    Richard S \textit{et al}. 2003 \textit{Phys. Rev. Lett.} \textbf{91} 010405
\end{thebibliography}
\end{document}